\documentclass[11pt]{article}
\usepackage[hmargin=3.5cm,vmargin=3cm]{geometry}
\usepackage{amsmath}
\usepackage{amssymb}
\usepackage{amsthm}
\usepackage[mathscr]{eucal}
\usepackage{enumitem}
\usepackage{bbm}
\usepackage{caption}
\usepackage{natbib}
\usepackage{graphicx}
\usepackage[pdftex]{color}
\definecolor{darkred}{rgb}{0.8,0.2,0.2}
\definecolor{darkblue}{rgb}{0.3,0.3,0.7}
\usepackage[pdftex,pdftitle={Turbocharging Monte Carlo pricing for the rough Bergomi model},pdfauthor={Ryan McCrickerd and Mikko S. Pakkanen},colorlinks=TRUE,allcolors=darkred]{hyperref}
\usepackage[text]{datetime}
\newdateformat{ukvardate}{%
\THEDAY{} \monthname[\THEMONTH] \THEYEAR}
\numberwithin{equation}{section}

\begin{document}
\title{Turbocharging Monte Carlo pricing \\ for the rough Bergomi model}
\date{
First version: 8 August 2017\\
This version: 16 March 2018}

\author{Ryan McCrickerd\thanks{
Department of Mathematics, 
Imperial College London, 
South Kensington Campus,
London SW7 2AZ, UK.
E-mail:\ 
\href{mailto:ryan.mccrickerd@jcrauk.com}{\nolinkurl{ryan.mccrickerd@jcrauk.com}}} \and
Mikko S. Pakkanen\thanks{
Department of Mathematics, 
Imperial College London, 
South Kensington Campus,
London SW7 2AZ, UK.
E-mail:\ 
\href{mailto:m.pakkanen@imperial.ac.uk}{\nolinkurl{m.pakkanen@imperial.ac.uk}}} \thanks{Corresponding author.}
}

\maketitle

\begin{abstract}

The rough Bergomi model, introduced by \cite{Bayer:2016}, is one of the recent rough volatility models that are consistent with the stylised fact of implied volatility surfaces being essentially time-invariant, and are able to capture the term structure of skew observed in equity markets. In the absence of analytical European option pricing methods for the model, we focus on reducing the runtime-adjusted variance of Monte Carlo implied volatilities, thereby contributing to the model's calibration by simulation. We employ a novel composition of variance reduction methods, immediately applicable to any conditionally log-normal stochastic volatility model. Assuming one targets implied volatility estimates with a given degree of confidence, thus calibration {\small RMSE}, the results we demonstrate equate to significant runtime reductions---roughly 20 times on average, across different correlation regimes.

\vspace*{1em}


\vspace*{1em}


\end{abstract}

\section{Background}

Rough volatility is a new paradigm in quantitative finance, motivated by the statistical analysis of realised volatility by \cite{Gatheral:2014} and the theoretical results on implied volatility by  \cite{Alos:2007} and \cite{Fukasawa:2011}. Rough volatility is generally characterised by the presence of a stochastic process \emph{rougher} that Brownian motion driving the volatility dynamics---fractional Brownian motion with Hurst exponent $H \in \left(0, \frac{1}{2}\right)$, popularised by \cite{Mandlebrot:1968}, is a convenient example of such a process. The rough Bergomi model (hereafter rBergomi) is the stochastic volatility pricing model developed by \cite{Bayer:2016}, which is consistent with the realised volatility model of \cite{Gatheral:2014} by means of an elegant change of measure. This \emph{rough stochastic volatility} pricing model outperforms classical counterparts by replicating implied volatility surface dynamics more accurately, being consistent with the stylised fact that the properties of volatility surfaces are essentially time-invariant, and by having fewer parameters---just three! The model is so named because of its relationship with the Bergomi variance curve model \citep{Bergomi:2005},
and may be seen as a non-Markovian generalisation of the latter. Due to the lack of Markovianity or affine structure, conventional analytical pricing methods, such as {\small PDEs} or Fourier transform, do not apply, motivating our quest for fast Monte Carlo pricing of vanilla instruments through a composition of variance reduction methods. While our focus is on the rBergomi model, our approach is applicable to a wide class of stochastic volatility models.

We work throughout on a filtered probability space $(\Omega,\mathcal{F},\{\mathcal{F}_t\}_{t\in\mathbb{R}},\mathbb{Q})$ that supports a two-dimensional Brownian motion $(W^1,W^2)$ with independent components, under the risk neutral measure $\mathbb{Q}$. The index $t$ will represent time in years from the present and we shall henceforth use the notation $\mathbb{E}[\cdot] = \mathbb{E}^{\mathbb{Q}}[\cdot|\mathcal{F}_0]$ unless we state otherwise. We let $S_t$ be an asset price process satisfying $\mathbb{E}[S_t] = 1$ for all $t\geq 0$, so define an \emph{out-of-the-money} ({\small OTM}) European call/put option with maturity $t$ and log-strike $k$ by its payoff,
\begin{equation}\label{eqPayoff}
(S_t - e^k)^+ := \max\big(w(S_t - e^k), 0\big),\quad w := -\mathbbm{1}_{(-\infty,0]}(k) + \mathbbm{1}_{(0,\infty)}(k),
\end{equation}
denoting its price observed today by $P(k,t)$.\footnote{We must stress the importance of this first step towards variance reduction. The implied volatilities generated when exclusively considering call or put option estimators are significantly noisier when they are respectively \emph{in-the-money}. This may be rationalised using the put-call parity, $\max\{S_t - e^k,0\} - \max\{e^k - S_t,0\} = S_t - e^k$. The methods we later employ remove this \emph{in-the-money variance}, but it is avoidable from the outset by always evaluating {\scriptsize OTM} options. Instead setting $w := \pm 1$ in \eqref{eqPayoff}, perceived variance reductions increase dramatically.} We define a Black--Scholes function $BS(\cdot)$ by
\begin{equation*}
BS(v;s,k) := w\big( s \mathcal{N}(wd_+) - e^k \mathcal{N}(wd_-) \big), \quad d_\pm := (\log s - k) /\sqrt{v} \pm \sqrt{v}/2,
\end{equation*}
where $\mathcal{N}(\cdot)$ represents the Gaussian cumulative distribution function.\footnote{This later enables use of the famed result $\log S \sim \mathcal{N}\left(\log s - \frac{1}{2}v,v\right) \implies \mathbb{E}[(S - e^k)^+] = BS(v;s,k)$. The somewhat unusual implied definition $k := \log K$, for strike $K$, compared with $k := \log(K/s)$, is used so $k$ remains fixed when we later vary $s$ through time.} The implied volatility $\sigma_{BS}(k,t)$ of an observed price $P(k,t)$ is thus defined using the relationship
\begin{equation*}
\sigma^2_{BS}(k,t)t = BS^{-1}(P(k,t); 1, k).
\end{equation*}

\subsection{The rBergomi model}

We adopt the rBergomi model \citep{Bayer:2016} for the price process $S_t$, and define it here by
\begin{equation}\label{eqrBergomi}
\begin{aligned}
S_t & = \mathcal{E}\left( \int_0^\cdot \sqrt{V_u} \mathrm{d} \left( \rho W^1_u + \sqrt{1 - \rho^2} W^2_u \right) \right)_t, \\ V_t & = \xi_0(t) \exp\left( \eta W^\alpha_t -\frac{\eta^2}{2}t^{2\alpha + 1}\right),
\end{aligned}
\end{equation}
where $\mathcal{E}(\cdot)$ denotes the stochastic exponential\footnote{Recall that for continuous semimartingale $X$, the stochastic exponential is defined $\mathcal{E}(X)_t:=\exp\left(X_t - X_0 - \frac{1}{2}[X]_t\right)$.} and $\eta>0$ and $\rho\in [-1,1]$ are parameters. We refer to $V_t$ as the variance process, and to $\xi_0(t) = \mathbb{E}[V_t] \in \mathcal{F}_0$ as the forward variance curve. In \eqref{eqrBergomi}, $W^\alpha$ is a certain Volterra process, also known as the \emph{Riemann-Liouville process}, defined by
\begin{equation*}
W_t^\alpha := \sqrt{2\alpha + 1} \int_0^t (t - u)^\alpha \mathrm{d} W^1_u
\end{equation*}
for $\alpha \in \left(-\frac{1}{2},0\right)$. This is a centred, locally $(\alpha + \frac{1}{2} - \epsilon)$-H\"{o}lder continuous, Gaussian process with  $\mathrm{Var}[W_t^\alpha] = t^{2\alpha + 1}$, and \emph{is not} a martingale, having negatively correlated increments, not even a semimartingale.

In order to simulate the process $W_t^\alpha$ efficiently and accurately, we utilise the first-order variant ($\kappa = 1$) of the \emph{hybrid scheme} \citep{Bennedsen:2015b}, which is based on the approximation
\begin{equation}\label{eqHybrid}
W_{\frac{i}{n}}^\alpha \approx \widetilde{W}_{\frac{i}{n}}^\alpha := \sqrt{2\alpha + 1} \left(\int^{\frac{i}{n}}_{\frac{i-1}{n}} \left(\frac{i}{n}-s\right)^\alpha \mathrm{d} W^1_u + \sum_{k=2}^i\left( \frac{b_k}{n} \right)^\alpha \Big(W^1_{\frac{i-(k-1)}{n}}-W^1_{\frac{i-k}{n}}\Big) \right),
\end{equation}
where
\begin{equation*}
b_k := \left(\frac{k^{\alpha+1}-(k-1)^{\alpha+1}}{\alpha+1}\right)^{\frac{1}{\alpha}}.
\end{equation*} 
Employing the fast Fourier transform to evaluate the sum in \eqref{eqHybrid}, which is a discrete convolution, a skeleton $\widetilde{W}_{0}^\alpha,\widetilde{W}_{\frac{1}{n}}^\alpha,\ldots,\widetilde{W}_{\frac{\lfloor nt \rfloor}{n}}^\alpha$ can be generated in $\mathcal{O}(n\log n)$ floating point operations.

\begin{figure}[t]
    \centering
    \includegraphics[width=0.425\linewidth]{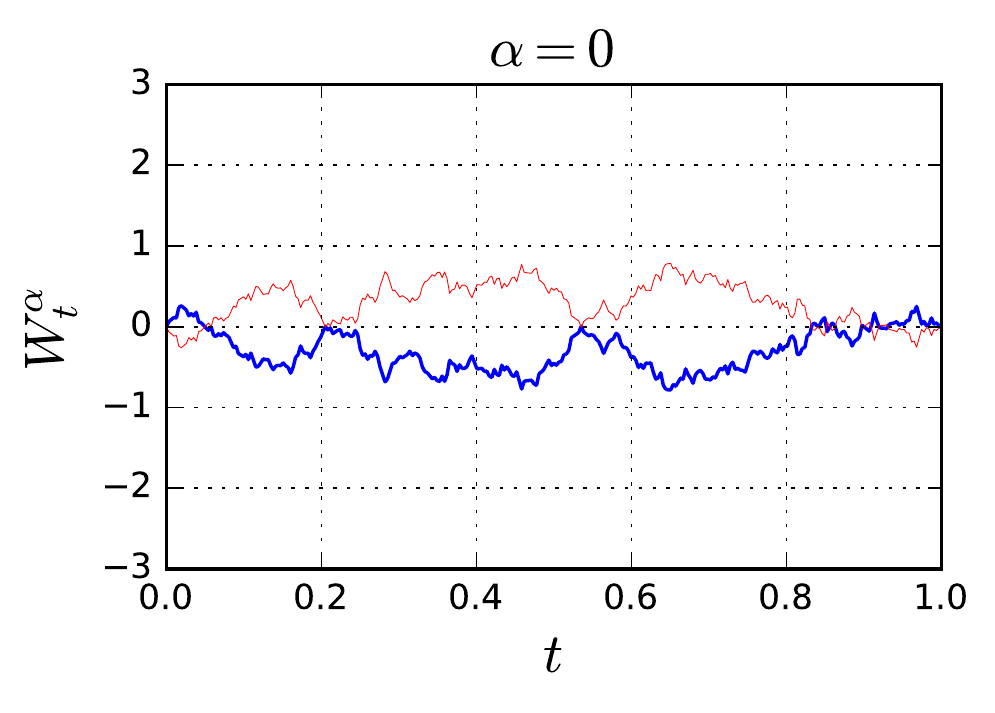}
    \includegraphics[width=0.425\linewidth]{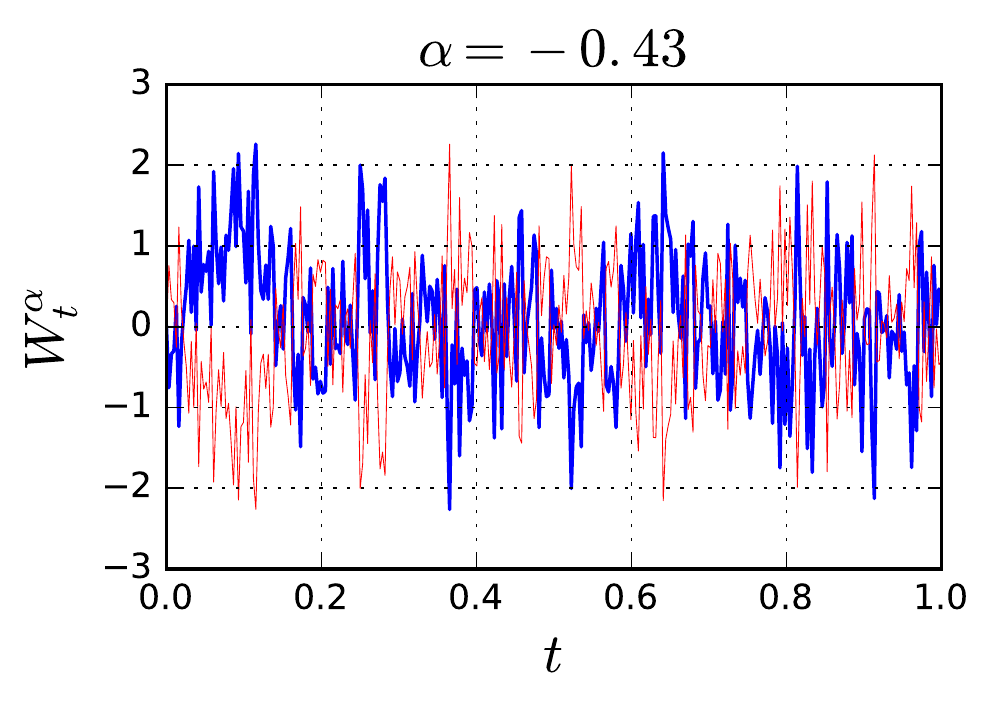}
    \caption{Sample paths of the Volterra process $W^\alpha$ for $\alpha = 0$, for which the process coincides with Brownian motion, and $\alpha = -0.43$. Each are $\mathcal{N}(0,t^{2\alpha + 1})$-distributed, so coincide at $t = 1$. A much greater short-time, \emph{i.e.} $t \ll 1$, variance is exhibited, however, when $\alpha = -0.43$. This \emph{explosive} short-time variance, generated when $\alpha$ is close to $-\frac{1}{2}$, leads to short-time implied volatilities observed in practice. We present here antithetic paths on a 312-point time grid.}\label{figVolterra}
\end{figure}

\begin{figure}[t]
    \centering
    \includegraphics[width=0.425\linewidth]{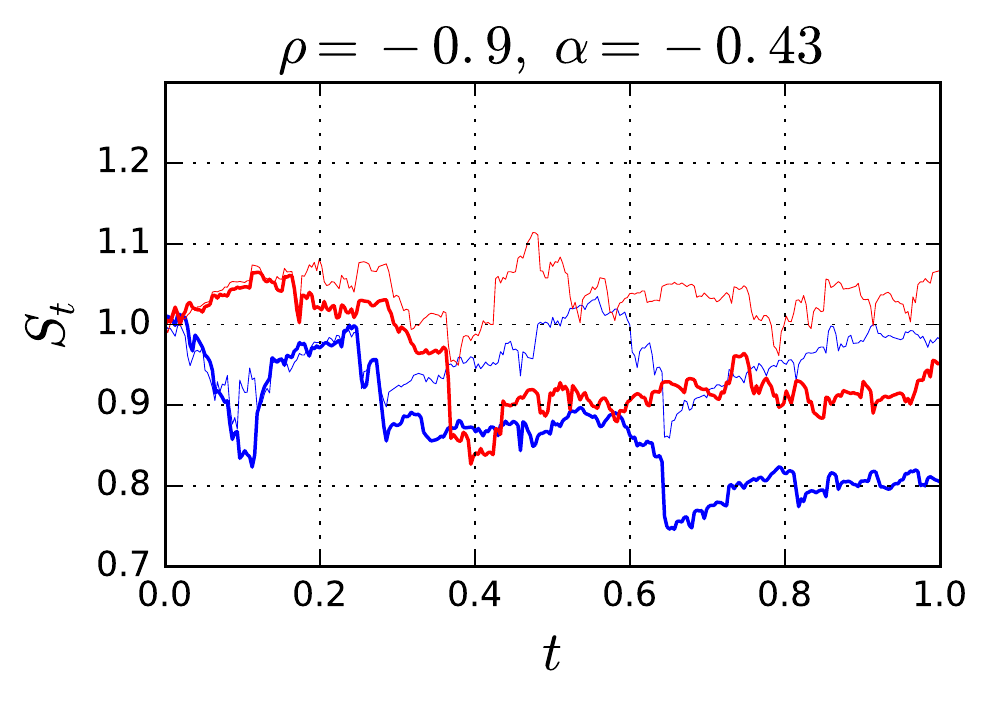}
    \includegraphics[width=0.425\linewidth]{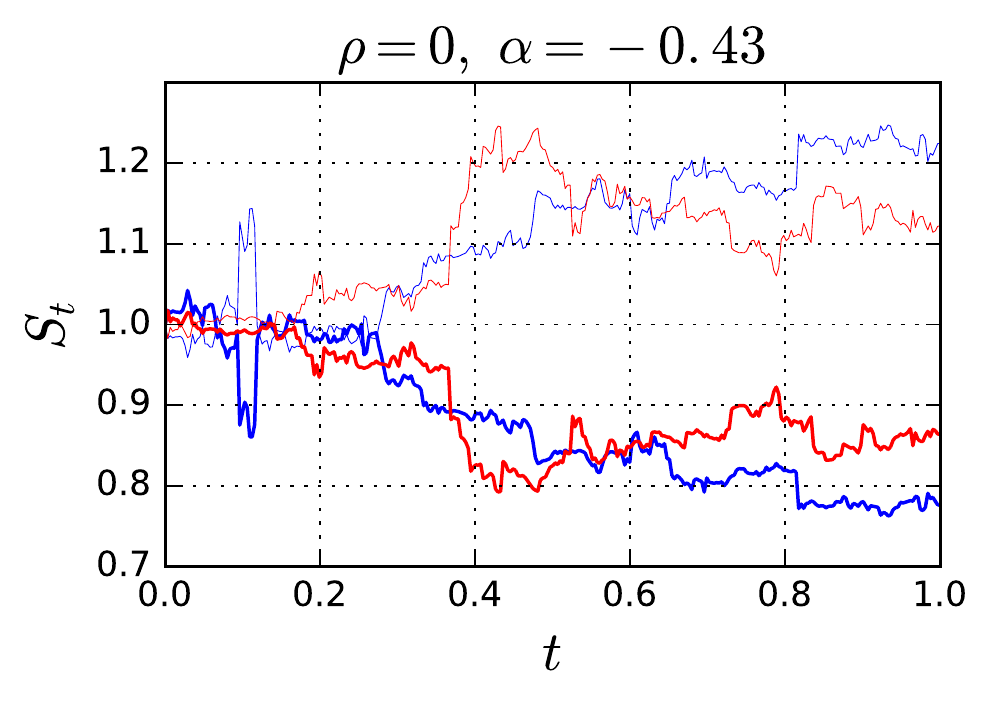}
    \vspace{-5mm}
    \caption{Sample rBergomi price paths using $\xi = 0.235^2$, $\eta = 1.9$, $\rho$ and $\alpha$ as stated. The price process, despite being a continuous martingale, exhibits jump-like behaviour when the Volterra, thus variance, process peaks. These price paths are based on antithetic paths of $(W^1,W^2)$, again on a 312-point time grid.}\label{figPrice}
\end{figure}

We demonstrate Volterra sample paths in Figure \ref{figVolterra}, which lead directly to the rBergomi price sample paths of Figure \ref{figPrice}.\footnote{We provide Python code on GitHub (\url{https://github.com/ryanmccrickerd/rough_bergomi}) and Jupyter notebooks that are able to reproduce sample paths and turbocharged implied volatilities.} The parameters of $\eta = 1.9$ and $\rho = -0.9$ there used are demonstrated by \cite{Bayer:2016} to be remarkably consistent with the \small SPX \normalsize market on 4 February 2010, and form the basis for our experiment, along with the case $\rho = 0$, which is more applicable, generally speaking, to other asset classes that deserve our interest, such as {\small FX}. We refrain from formally naming these model parameters, but those seeking an intuitive understanding of their influence over implied volatilities might like \emph{smile} for $\eta$, \emph{skew} for $\rho$, and \emph{explosion} (of smile and skew) for $\alpha$.

\section{Implied volatility estimators}

\begin{figure}[t]
    \centering
    \begin{tabular}{@{}r@{}r@{}}
    \includegraphics[width=0.425\linewidth]{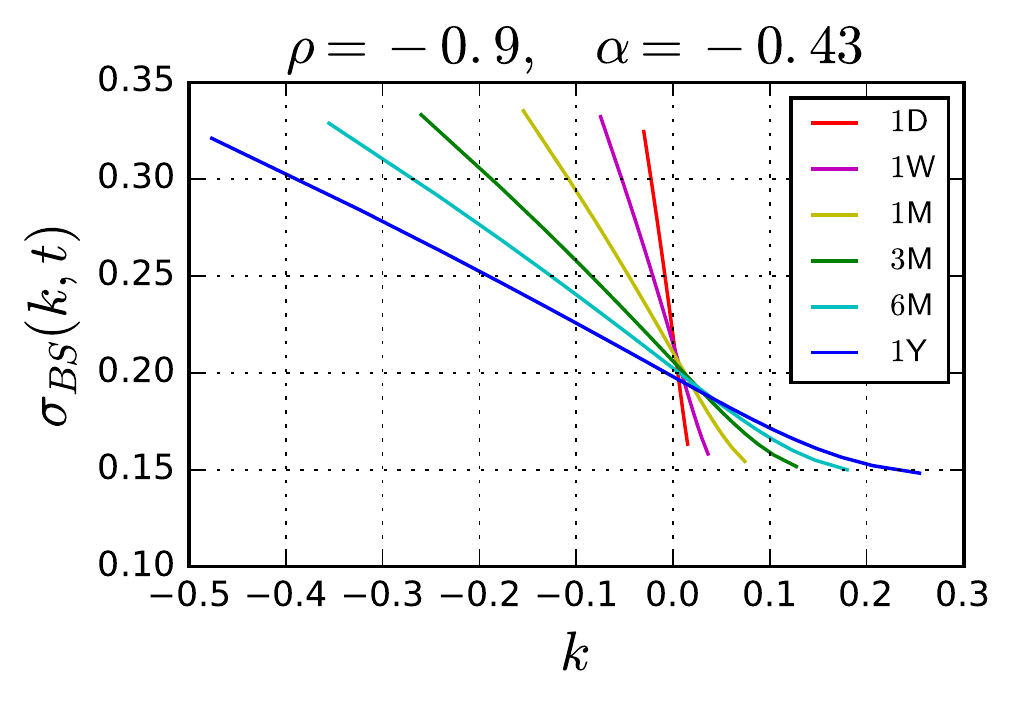} &
    \includegraphics[width=0.425\linewidth]{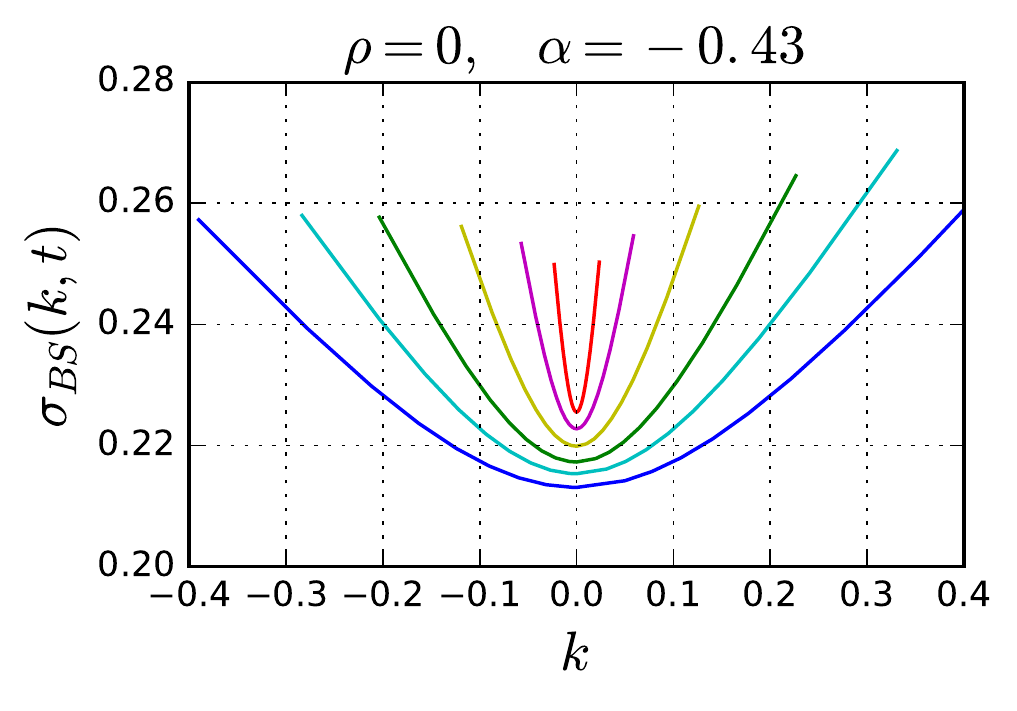} \\
    \includegraphics[width=0.425\linewidth]{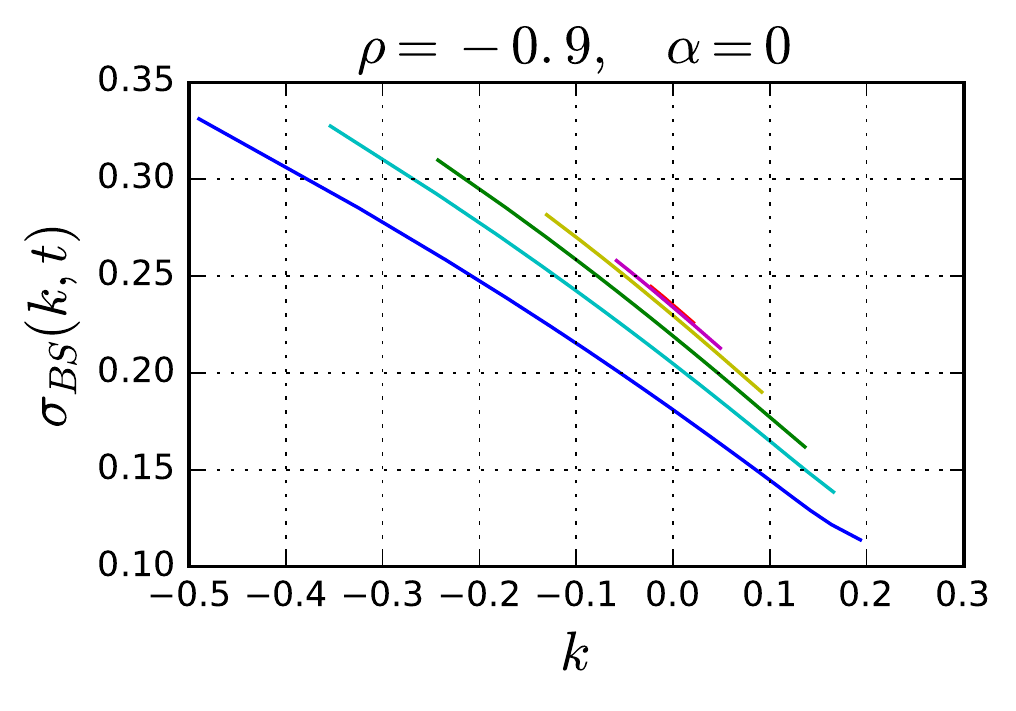} &
    \includegraphics[width=0.425\linewidth]{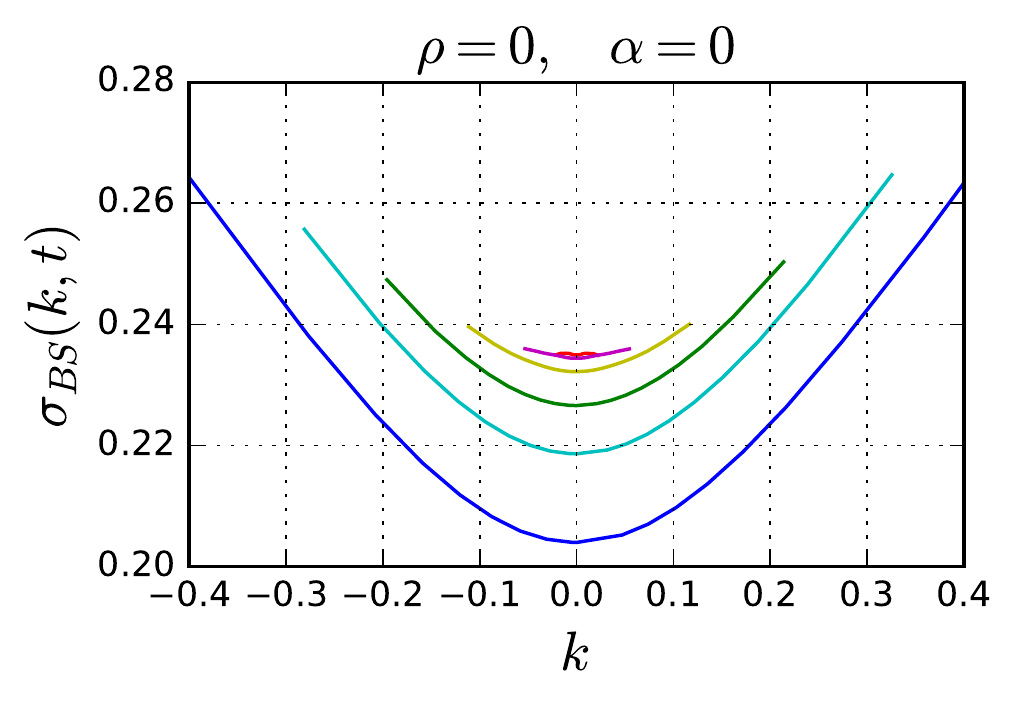}
    \end{tabular}
    \caption{Implied volatilities from a simulation of \eqref{eqrBergomi} for maturities ranging from one day to one year, using the Base estimator defined by \eqref{eqBaseEstimator}. Parameter values are $\xi_0(t) = \xi = 0.235^2$ and $\eta = 1.9$, with $\rho$ and $\alpha$ as stated. Log-strikes represent a range from 5 delta puts ($\mathcal{N}(-d_+) = 0.05$) to 5 delta calls ($\mathcal{N}(d_+) = 0.05$) in 5 delta increments (19 in total for each maturity). In the simulation, $400{,}000$ antithetic paths are used on $(W^1,W^2)$ and each maturity is separately discretised on a 312-point grid.}\label{figSmiles}
\end{figure}

Accepting the representation $P(k,t) = \mathbb{E}[(S_t - e^k)^+]$ of {\small OTM} option prices, we proceed to consider price estimators $\hat{P}_n(k,t)$ of the following form under the rBergomi model
\begin{equation}\label{eqEstimators}
\hat{P}_n(k,t) := \frac{1}{n}\sum_{i=1}^n \left( X_i + \hat{\alpha}_n Y_i \right) - \hat{\alpha}_n \mathbb{E}[Y], \quad \hat{\sigma}^n_{BS}(k,t)^2 t = BS^{-1}\big(\hat{P}_n(k,t);1,k\big),
\end{equation}
from which we derive implied volatility estimators $\hat{\sigma}^n_{BS}(k,t)$. Notice that these are always biased by the non-linearity of $BS(\cdot)$ and the requirement to take a square root.\footnote{We later report some bias, but we find that even when using $n = 1{,}000$, it is never practically meaningful.} In \eqref{eqEstimators}, $X_i$ and $Y_i$ are samples of random variables to be specified. For example, our \emph{Base} estimator shall be defined naturally by setting
\begin{equation}\label{eqBaseEstimator}
X = (S_t - e^k)^+,\quad Y = 0.
\end{equation}
A rich variety of implied volatility smiles generated using this estimator are presented in Figure \ref{figSmiles}, which will further aid intuition for this model. The case $\alpha = 0$ is comparable to classical stochastic volatility models in the absence of time-dependent or randomised parameters, or jump processes. Some admirable recent efforts in the randomised case are \cite{Mechkov:2016} and \cite{Jacquier:2017}, and for jumps \cite{Mechkov:2015}. On the contrary, when $\alpha = -0.43$, the explosions of skew and smile as $t \to 0$ are precisely as observed in practice.

In pursuit of a variance reducing estimator of the form \eqref{eqEstimators}, following \cite{Romano:1997}, we consider the orthogonal separation of the rBergomi price process $S_t$ into $S^1_t$ and $S^2_t$, where
\begin{equation*}
S^1_t := \mathcal{E}\left( \rho \int_0^\cdot \sqrt{V_u} \mathrm{d} W^1_u \right)_t, \quad S^2_t := \mathcal{E}\left( \sqrt{1 - \rho^2} \int_0^\cdot \sqrt{V_u} \mathrm{d} W^2_u \right)_t,
\end{equation*}
which allows us to capitalise on its conditional log-normality.\footnote{For a conditionally Gaussian process, our methods could be adapted using, for example, Hull--White price evaluation in place of Black--Scholes.} By conditional log-normality, we explicitly mean
\begin{equation}\label{eqCondLN}
\log S_t \ | \ \mathcal{F}^1_t \vee \mathcal{F}^2_0 \sim \mathcal{N} \left( \log S^1_t - \frac{1}{2}\left( 1 - \rho^2 \right) \int_0^t V_u \mathrm{d}u,\ \left( 1 - \rho^2 \right) \int_0^t V_u \mathrm{d}u \right),
\end{equation}
where we use natural filtrations $\mathcal{F}^i_t = \sigma\{W^i_u:u \leq t\}$, $i = 1,2$. Since both $\int_0^t V_u \mathrm{d}u$ and $S_t^1$ are measurable with respect to $\mathcal{F}_t^1$, this representation becomes intuitively clear when we imagine $S^1_t$ as a spot price, and $\left( 1 - \rho^2 \right) \int_0^t V_u \mathrm{d}u$ as the integrated variance originating from $S_t^2$, as is described by \cite{Romano:1997} and \cite{Bergomi:2016} in wider stochastic volatility frameworks. This separation facilitates our \emph{Mixed} estimator, which we define using \eqref{eqEstimators} with
\begin{equation}\label{eqMixtureEstimator}
X = BS\left( \left( 1-\rho^2 \right) \int_0^t V_u \mathrm{d}u; S_t^1, k \right),\quad Y = BS\left( \rho^2\left(\hat{Q}_n - \int_0^t V_u \mathrm{d}u\right); S^1_t, k \right),
\end{equation}
where the estimated parameters $\hat{\alpha}_n$ and $\hat{Q}_n$ will be soon made explicit.

The Mixed estimator represents the composition of the conditional Monte Carlo method with a control variate, which we have found to be individually most effective in the regimes $\rho = 0$ and $\rho = -0.9$ respectively. This use of $X$ represents the simulation of a conditional expectation because, following \eqref{eqCondLN}, we have the representation
\begin{equation*}
X = \mathbb{E}\big[( S_t - e^k)^+ \,\big|\,  \mathcal{F}^1_t \vee \mathcal{F}^2_0 \big].
\end{equation*}
The Tower property then ensures $\mathbb{E}[X]$ agrees with the expectation of the Base estimator. Amazingly, this eliminates all dependence on $W^2$, and in theory guarantees a variance reduction. The component $Y$ in the Mixed estimator admits a representation as the time $t$ price of a \emph{Timer} option with variance budget $\rho^2\hat{Q}_n$, written on the parallel component $S^1_t$ of the price process.\footnote{That analytical Timer option prices should be available under stochastic volatility models is intuitively clear, but a probabilistic interpretation of why is wonderful: $\log S_t + \frac{1}{2}\int_0^t V_u\mathrm{d}u = \int_0^t \sqrt{V_u}\mathrm{d}W^1_u$ is a continuous local martingale starting at zero on $(\Omega,\mathcal{F},\{\mathcal{F}_t\}_{t\geq 0},\mathbb{Q})$, so defining the stopping time $\tau_Q := \inf\{u>0:\int_0^u V_s \mathrm{d}s = Q\}$, the Dubins--Schwarz theorem provides $B_Q := \log S_{\tau_Q} + \frac{1}{2}Q$ is a Brownian motion on $(\Omega,\mathcal{F},\{\mathcal{F}_{\tau_Q}\}_{Q\geq 0},\mathbb{Q})$.} The process $Y = Y_t$ is clearly a martingale, because it has the representation 
\begin{equation*}
Y_t = \mathbb{E}\big[(S^1_{\tau_{\hat{Q}_n}} - e^k)^+\,\big|\,\mathcal{F}^1_t\big],\quad \tau_{\hat{Q}_n} := \inf\left\{u>0:\int_0^u V_s \mathrm{d}s = \hat{Q}_n\right\},
\end{equation*}
as is the case for any tradeable asset. For all maturities $t$ we are therefore able to make use of the following expectation in \eqref{eqEstimators},
\begin{equation*}
\mathbb{E}[Y] = \mathbb{E}[Y_0] = BS\big( \rho^2\hat{Q}_n ; 1, k \big).
\end{equation*}

We compute $\hat{\alpha}_n$ and $\hat{Q}_n$ post-simulation from sampled $X_i$, $Y_i$ and $\left(\int_0^t V_u \mathrm{d}u\right)_i$, using
\begin{equation}\label{eqParamDefs}
\hat{\alpha}_n := -\frac{\sum_{i=1}^n \left(X_i - \bar{X}_n \right)\left(Y_i - \bar{Y}_n \right)}{\sum_{i=1}^n \left(Y_i - \bar{Y}_n \right)^2},\quad \hat{Q}_n := \sup\left\{ \left(\int_0^t V_u \mathrm{d}u\right)_i : i=1,\dots,n \right\},
\end{equation}
meaning that our variance reducing methods lose their relationship with hedging strategies in practice. The former is known to asymptotically minimise the variance of $\hat{P}_n(k,t)$ for any control variate, see for example \citet[pp.\ 138--139]{Asmussen:2007}. The choice of $\hat{Q}_n$ might seem unnerving, but is the minimum that avoids the computation of stopping times when evaluating $Y$, which we find to be relatively computationally expensive.\footnote{For example, one might set $Y = BS \left( Q - \int_0^{t\wedge\tau_Q} V_u \mathrm{d}u; S_{t\wedge\tau_Q}, k \right)$, with $\tau_Q$ as defined above.} The choice otherwise ensures that $Y$ outperforms the more obvious martingale control variate $wS^1_t$, effectively because the following limit holds\footnote{It is worth appreciating that in the seemingly awkward limits of $\rho \to 0$ and $\rho \to \pm 1$, the Mixed estimator performs like the conditional Monte Carlo method and a control variate \emph{independently}, respectively, by design.}
\begin{equation*}
\lim_{Q \to \infty} Y = \lim_{Q \to \infty} BS\left( \rho^2\left(Q - \int_0^t V_u \mathrm{d}u\right); S^1_t, k \right) = wS^1_t.
\end{equation*}

Finally, we briefly explain our use of antithetic sampling for the Mixed estimator. We draw a path of $W^1$ over the interval $[0,t]$, and appeal to the symmetry in distribution of $S^{1,\pm}_t$, defined by
\begin{align*}
S^{1,\pm}_t & = \mathcal{E}\left\{\pm \rho \int_0^t \sqrt{V^\pm_u} \mathrm{d} W^1_u \right\}, & V^\pm_t & = \xi_0(t)\exp\left(-\frac{\eta^2}{2}t^{2\alpha + 1}\right)\left(V^\circ_t\right)^{\pm 1}, \\
 V^\circ_t &=  \exp\left( \eta W^\alpha_t \right).
\end{align*}
Notice that, besides providing an outright variance reduction, this immediately halves the number of required Volterra paths, reducing total runtime significantly. Now that the Mixed estimator is fully defined, we summarise the estimators from which it was developed in Table \ref{tabEstimators}. The Conditional estimator and some methods related to our Controlled estimator, for example, the \emph{Timer option-like algorithm}, may be found in \citet[pp.\ 336--342]{Bergomi:2016} in a general stochastic volatility setting.

\begin{table}[h]
\centering
\begin{tabular}{|l|l|c|l|c|}
\cline{1-1} \cline{3-3} \cline{5-5}
Estimator &  & $X$ &  & $Y$ \\ \cline{1-1} \cline{3-3} \cline{5-5} 
Base &  & $\left(S_t - e^k\right)^+$ &  & $0$ \\
Conditional &  & $BS\big((1 - \rho^2) \int_0^t V_u \mathrm{d}u; S^1_t, k\big)$ &  & $0$ \\
Controlled &  & $(S_t - e^k)^+$ &  & $BS\big(\hat{Q}_n - \int_0^t V_u \mathrm{d}u;S_t, k\big)$ \\
Mixed &  & $BS\big(( 1-\rho^2 ) \int_0^t V_u \mathrm{d}u; S_t^1, k \big)$ &  & $BS\big( \rho^2 \big(\hat{Q}_n - \int_0^t V_u \mathrm{d}u \big); S^1_t, k \big)$ \\ \cline{1-1} \cline{3-3} \cline{5-5} 
\end{tabular}
\caption{Intermediate estimator definitions, for use in \eqref{eqEstimators}, which lead to the Mixed estimator. The quantities $\hat{\alpha}_n$ and $\hat{Q}_n$ are always defined generally by \eqref{eqParamDefs}, when $Y \neq 0$. Considering the Conditional and Controlled estimators, the Mixed estimator is clearly their natural extension, which tends to each in the limits $\rho \to 0$ and $\rho \to \pm 1$ respectively.}\label{tabEstimators}
\end{table}

In the next section, we conduct an experiment to compare implied volatilities derived from our Base and Mixed estimators. We use a relatively low number of paths, comparing resulting bias and variances with the higher quality data in Figure \ref{figSmiles}.\footnote{Number of paths is almost arbitrary, because we find our estimators adhere neatly to the scaling properties implied by the central limit theorem: to halve observed standard deviations, simply quadruple number of paths.} Following this comparison, we proceed to briefly demonstrate the impact of our work on the rBergomi parameters driving smile and skew, $\eta$ and $\rho$, in an experiment assessing the calibration accuracy of those parameters by simulation. All of this is implemented in Python, although we use the NumPy library heavily to ensure C{\small ++}-like runtimes. We use the default NumPy pseudo-random number generator (Mersenne Twister). The performance of all implied volatility estimators can be improved slightly by instead using quasi-random numbers (low-discrepancy sequences, e.g., Sobol), but our experiments with Sobol sequences, obtained using the Sobol Julia module, suggest that the improvement is not dramatic. With a focus on results, practical application and building intuition for the rBergomi model, we simply summarise results for the intermediate estimators, and point to \cite{Asmussen:2007} for some general theory underlying this work.

\section{Variance reduction}

As is widely understood by practitioners of Monte Carlo methods, \emph{the greatest gains from variance reduction techniques result from exploiting specific features of the problem at hand}---adapted from \cite{Glasserman:2004}. Although the theory of antithetic sampling, conditional Monte Carlo, and control variates are well understood, these methods are somewhat meaningless without refinement to our estimation of implied volatilities under the rBergomi model. 

\subsection{Experiment design}

We now fix the maturity $t = 0.25$, so may drop its reference, and rBergomi parameters $\xi = 0.235^2$, $\eta = 1.9$ and $\alpha = -0.43$. We consider the two correlation regimes of $\rho = -0.9$ and $\rho = 0$, and three log-strikes representing 10 delta put, {\small ATM}, and 10 delta call options in each regime.\footnote{Specifically, this means $\mathcal{N}(-d_+) = 0.10$, $k = 0$ and $\mathcal{N}(d_+) = 0.10$ respectively.} We consider sampling $\hat{\sigma}^n_{BS}(k)$ from \eqref{eqEstimators} $N$ times, in order to obtain a sequence $\{\hat{\sigma}^{n}_{BS}(k)_i\}_{i=1}^{N}$ of estimates. Given the following central limit theorem for estimated prices
\begin{equation*}
\sqrt{n}\big(\hat{P}^n(k,t) - P(k,t)\big) \xrightarrow[n\to\infty]{D} \mathcal{N}\left(0, v_\infty\right),
\end{equation*}
with $v_\infty: = \lim_{n\to\infty}\mathrm{Var}[X + \hat{\alpha}_n Y]$ and $X$,\ $Y$ as in \eqref{eqEstimators}, the Delta method provides the additional convergence  
\begin{equation*}
\sqrt{n}\big(\hat{\sigma}^n_{BS}(k) - \sigma_{BS}(k)\big) \xrightarrow[n\to\infty]{D} \mathcal{N}\big(0, v_\infty \left(2t \sigma_{BS}(k,t)BS'(\sigma_{BS}^2(k,t)t;1,k) \right)^{-2}\big).
\end{equation*}
Fixing $n = 1{,}000$ and $N = 1{,}000$, we therefore plot histograms of the sampled sequences $\{\hat{\sigma}^n_{BS}(k)_i - \sigma_{BS}(k)\}_{i=1}^N$ alongside fitted normal distributions. Of course, we don't truly know $\sigma_{BS}(k)$, hence the use of the results in Figure \ref{figSmiles} as proxies. They are provided for the relevant {\small 3M} maturity in the following table for clarity. 

\begin{table}[t]
\centering
\begin{tabular}{|l|rrr|l|l|rrr|}
\cline{1-4} \cline{6-9}
$\rho = -0.9$ & \multicolumn{1}{c}{10P} & \multicolumn{1}{c}{ATM} & \multicolumn{1}{c|}{10C} &  & $\rho = 0$ & \multicolumn{1}{c}{10P} & \multicolumn{1}{c}{ATM} & \multicolumn{1}{c|}{10C} \\ \cline{1-4} \cline{6-9} 
$k$ & $-0.1787$ & 0.0000 & 0.1041 &  & $k$ & $-0.1475$ & 0.0000 & 0.1656 \\
$\sigma_{BS}(k)$ & 29.61 & 20.61 & 15.76 &  & $\sigma_{BS}(k)$ & 24.17 & 21.73 & 24.66 \\ \cline{1-4} \cline{6-9} 
\end{tabular}\caption{Log-strikes and implied volatilities (scaled by 100) selected from the top two plots of Figure \ref{figSmiles} respectively. The reproduction of these results by the Base and Mixed estimators shall form the basis of our experiment.}\label{tabLogStrikes}
\end{table}

In order to compare estimators in a manner which is both runtime-adjusted and weakly dependent on the choice of $n$, we take guidance from \cite{Glasserman:2004} when defining our measure of $\hat{\sigma}^n_{BS}(k)$ variance. To this end, we let $\tau$ denote the runtime in milliseconds to produce a single $\hat{\sigma}^n_{BS}(k)_i$ estimation.\footnote{Given the target application of this work, we must approximate a runtime which is indicative of the time taken by a minimisation routine of implied volatility {\scriptsize RMSE}s. This in itself is ambiguous, given, amongst other things, this time will be affected by which of the rBergomi parameters are being calibrated. Specifically, we let $\tau$ be the time to produce the sample $\{\hat{\sigma}^n_{BS}(k)_i - \sigma_{BS}(k)\}_{i=1}^N$, divided by $N$.} Considering log-strikes $\{k_i\}_{i = 1}^m$, we thus define the \emph{mean squared error} and \emph{mean runtime-adjusted squared error} measures of our estimators respectively by 
\begin{equation}\label{eqMeasureDef}
\phi^2 := \frac{1}{m} \sum_{i=1}^{m} \hat{\sigma}^2_{n,N,k_i},\quad \psi^2 := \frac{\tau}{m} \sum_{i=1}^{m} \hat{\sigma}^2_{n,N,k_i},
\end{equation}
where we simply estimate
\begin{equation*}
\hat{\sigma}^2_{n,N,k} := \frac{1}{N - 1}\sum_{i=1}^N \left(\hat{\sigma}^n_{BS}(k)_i - \sigma_{BS}(k) \right)^2.
\end{equation*}
Notice that $\psi^2$ is in theory asymptotically independent of $n$, since $\tau$ scales like $n$ and for each $k_i$, $\hat{\sigma}^2_{n,N,k_i}$ scales asymptotically like $1/n$. Having fixed $n$ and $N$, for ease of computations, we may therefore use ratios of estimator $\psi^2$ values in order to approximate the relative runtime to achieve a fixed $\phi$ value (corresponding to a calibration {\small RMSE}), since $\tau = \psi^2/\phi^2$. These observations are reflected in practice, certifying $\psi^2$ as a sensible means for comparison. We stress that our use of $n = 1{,}000$ is only for convenience, and to demonstrate the performance of the Mixed estimator with such few paths. Indeed, because we find that all estimators' standard deviations adhere to the scaling suggested by the central limit theorem, one may predictably shrink observed confidence intervals by increasing $n$.

\subsection{Results}

\begin{figure}[!t]
    \centering
    \begin{tabular}{@{}r@{}}
    \includegraphics[width=0.9\linewidth]{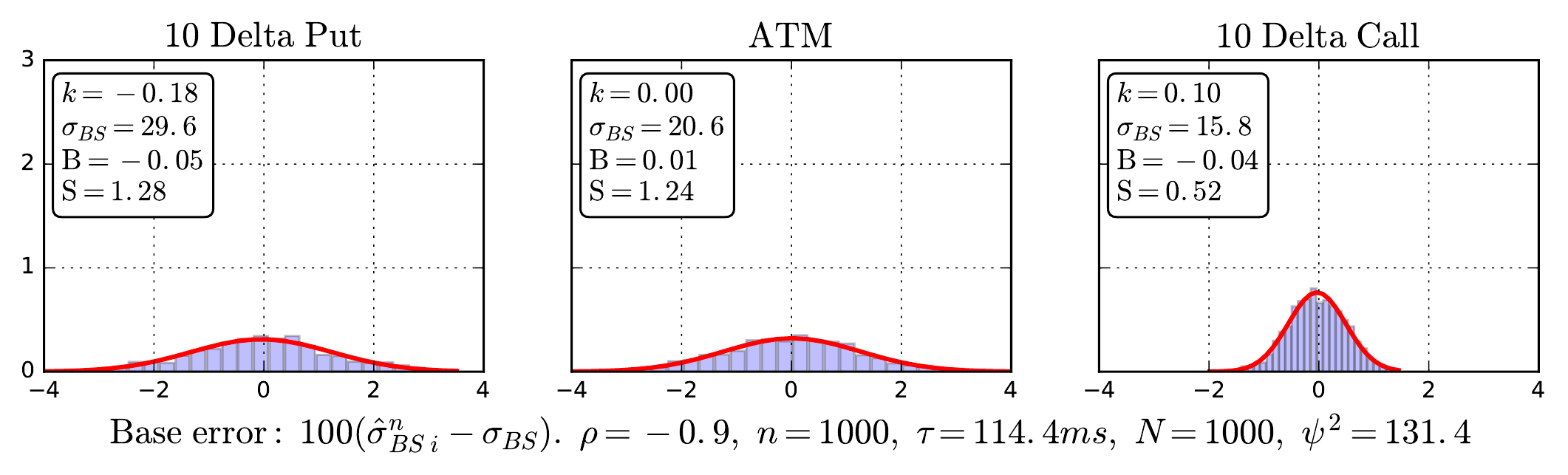}\\
     \includegraphics[width=0.9\linewidth]{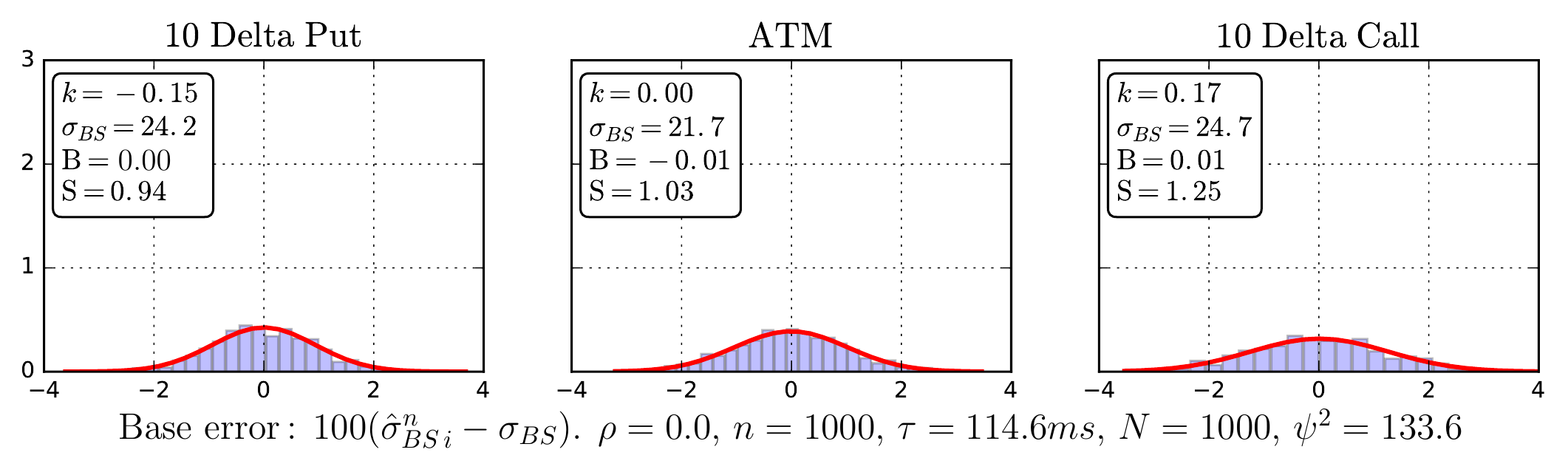}
     \end{tabular}
    \vspace{-5mm}
    \caption{Implied volatility estimator $\hat{\sigma}^n_{BS}(k)$ distributions using the Base estimator of \eqref{eqBaseEstimator}, for $\rho = -0.9$ (top) and $\rho = 0$ (bottom). Individual bias and standard deviations of each sample $\{\hat{\sigma}^{n}_{BS}(k)_i\}_{i=1}^{N}$ are labelled B and S. Average runtimes $\tau$ and runtime-adjusted squared errors, $\psi^2$ defined in \eqref{eqMeasureDef}, are also shown. These, like all that follow, were recorded by a laptop running macOS Sierra 10.12 with a 2 GHz Intel Core i5 processor and 8 GB memory.}\label{figBaseHist}
    \includegraphics[width=0.425\linewidth]{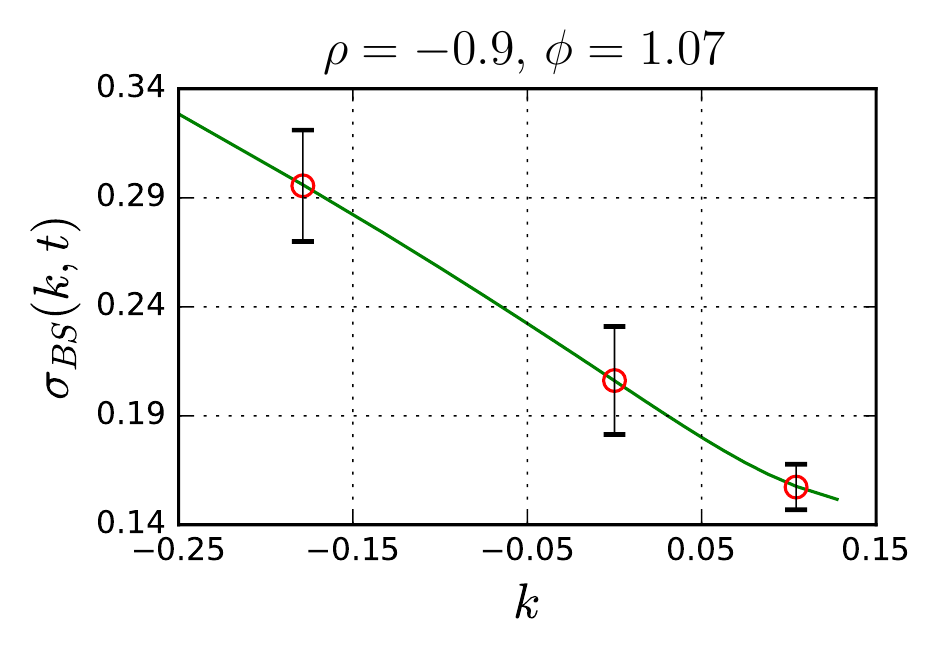}
    \includegraphics[width=0.425\linewidth]{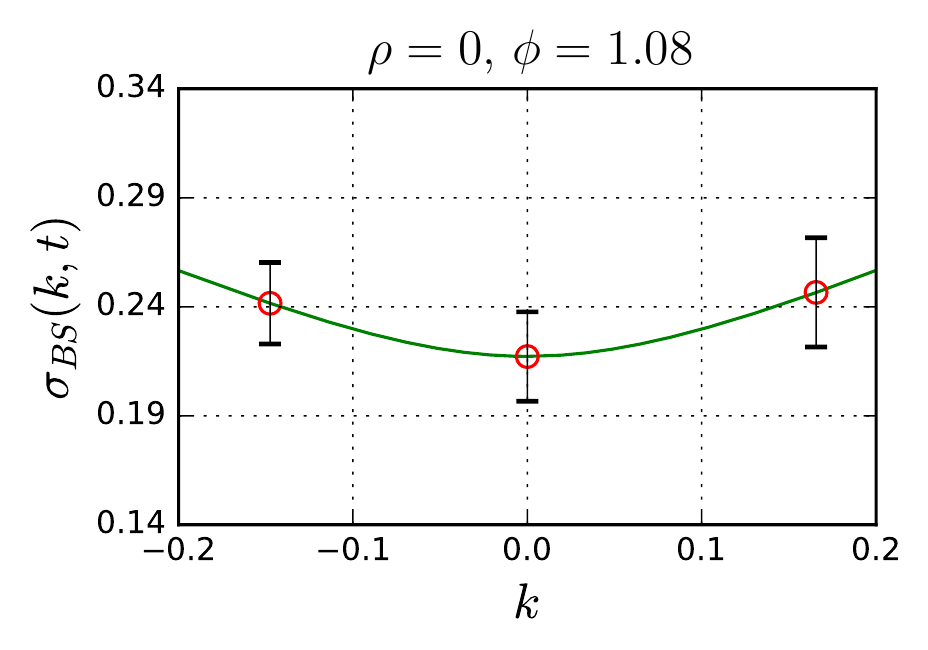}
    \vspace{-5mm}
    \caption{Implied volatility estimator $\hat{\sigma}^n_{BS}(k)$ expectations (red), with 95\% confidence intervals (black), using the Base estimator of \eqref{eqBaseEstimator}. Each $\hat{\sigma}^n_{BS}(k)$ is sampled $N = 1{,}000$ times, using $n = 1{,}000$ paths.}\label{figRMSEsBase}
\end{figure}

\begin{figure}[!t]
    \centering
     \begin{tabular}{@{}r@{}}
   \includegraphics[width=0.9\linewidth]{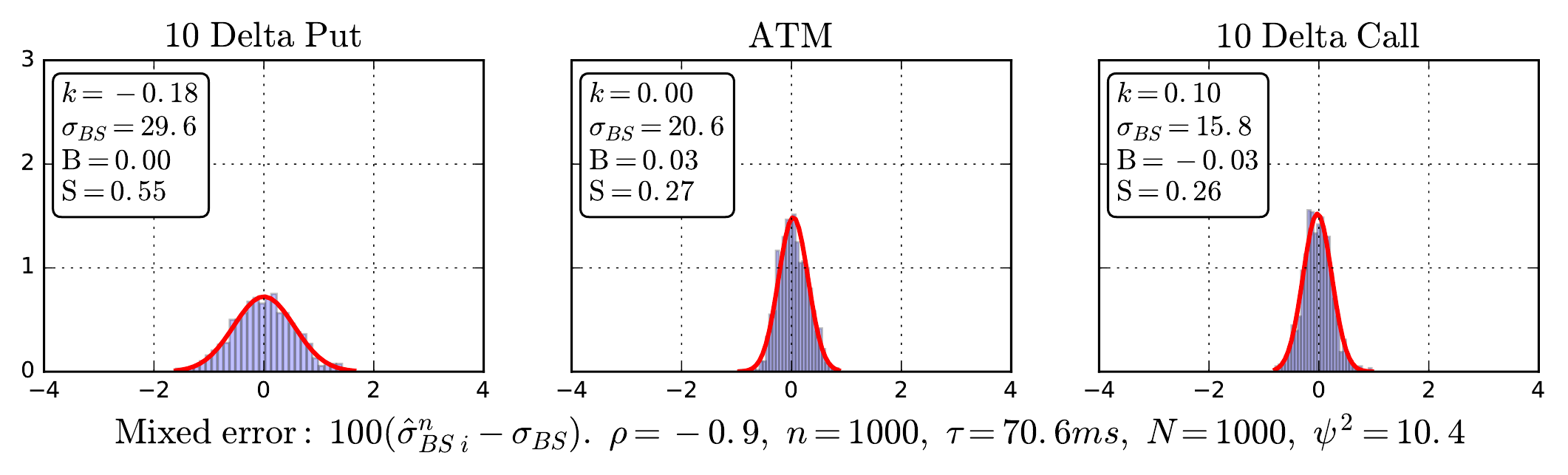} \\
    \includegraphics[width=0.9\linewidth]{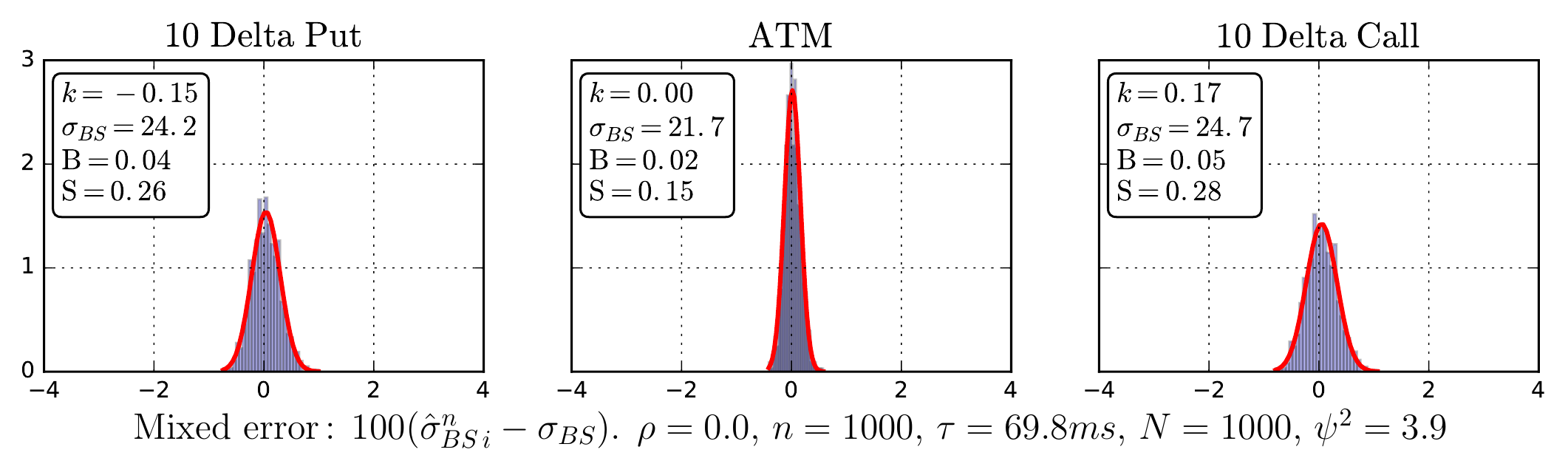}
 \end{tabular}
    \vspace{-5mm}
    \caption{Implied volatility estimator $\hat{\sigma}^n_{BS}(k)$ distributions using the Mixed estimator of \eqref{eqMixtureEstimator}, for $\rho = -0.9$ (top) and $\rho = 0$ (bottom). Individual bias and standard deviations of each sample $\{\hat{\sigma}^{n}_{BS}(k)_i\}_{i=1}^{N}$ are labelled B and S. Average runtimes $\tau$ and runtime-adjusted squared errors, $\psi^2$ defined in \eqref{eqMeasureDef}, are also shown.}\label{figMixedHist}
%
    \includegraphics[width=0.425\linewidth]{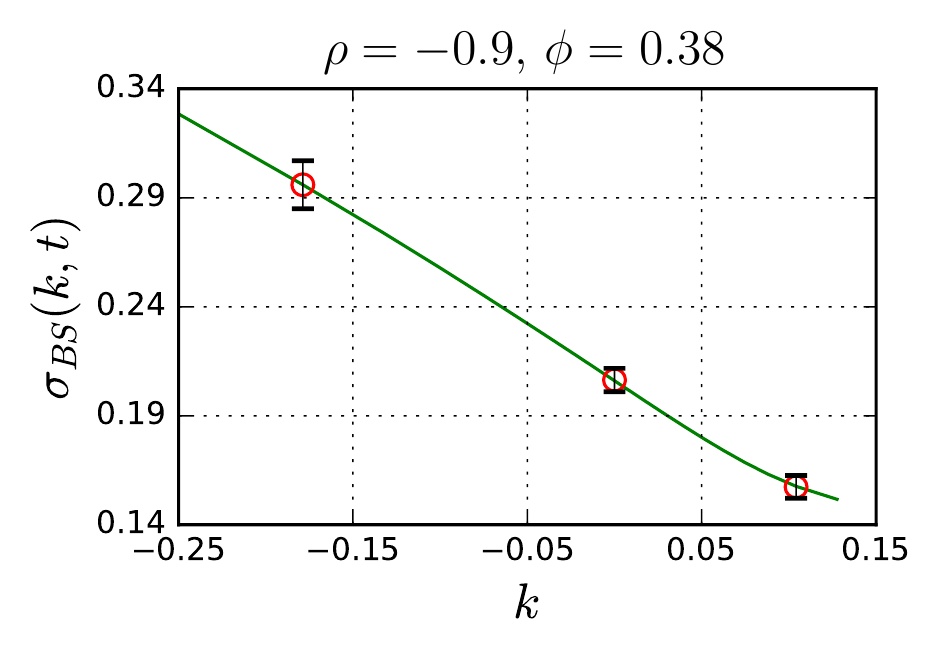}
    \includegraphics[width=0.425\linewidth]{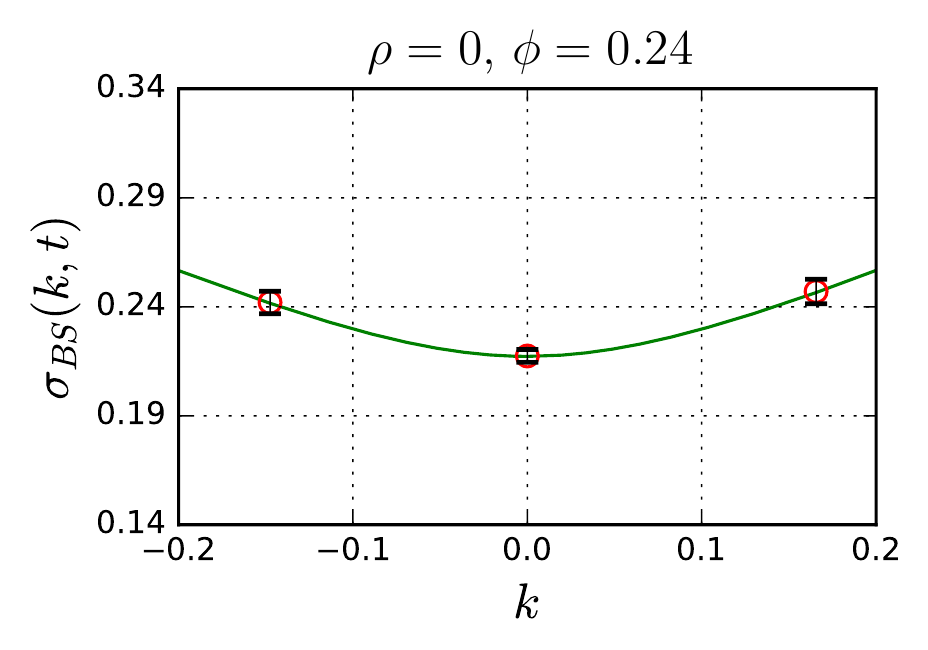}
    \vspace{-5mm}
    \caption{Implied volatility estimator expectations, with 95\% confidence intervals, using the Mixed estimator of \eqref{eqMixtureEstimator}, alongside data from Figure \ref{figSmiles}. Each $\hat{\sigma}^n_{BS}(k)$ is sampled $N = 1{,}000$ times, using $n = 1{,}000$ paths.}\label{figRMSEsMixed}
\end{figure}

Histograms with fitted normal distributions, and implied volatility confidence intervals are shown for the Base estimator in Figures \ref{figBaseHist} and \ref{figRMSEsBase}, and for the Mixed estimator in Figures \ref{figMixedHist} and \ref{figRMSEsMixed}, respectively. Histograms are labelled with each applicable log-strike $k$, target \emph{implied volatility} $\sigma_{BS}$ and \emph{bias} (B) and \emph{standard deviation} (S) of the sample $\{\hat{\sigma}^{n}_{BS}(k)_i\}_{i=1}^{N}$. We stress again that $\psi^2$ is the measure that should be used to determine relative estimator runtimes to achieve a given implied volatility confidence interval. 

The Base estimator results in Figure \ref{figBaseHist} demonstrate standard deviations, around 1 percentage point (i.e., one Vega), which render it unfit for practical purposes. This, of course, is not surprising when using just $n = 1{,}000$ paths, and these results are nevertheless important for aiding comparison. Figure \ref{figRMSEsBase} places these results and 95\% confidence intervals over the equivalent implied volatilities from Figure \ref{figSmiles}, also showing root mean squared errors, $\phi$. In general, one finds greatest variances at the 10 delta call strike, but in the case of $\rho = -0.9$, this effect is dominated by the price process inheriting greater variances for low strikes.

The Mixed estimator results in Figure \ref{figMixedHist} demonstrate standard deviations much lower than 1 percentage point (i.e., one Vega). Even in the most uncertain case, the sampled implied volatility is within 1.1 percentage points of the known value 29.6\%, 95\% of the time. We consider this remarkable, evidently, considering the number of paths, $n = 1{,}000$, used. Figure \ref{figRMSEsBase} places these results and 95\% confidence intervals over the equivalent implied volatilities from Figure \ref{figSmiles}, also showing root mean squared errors, $\phi$.

\begin{figure}[h]
    \centering
    \includegraphics[width=0.425\linewidth]{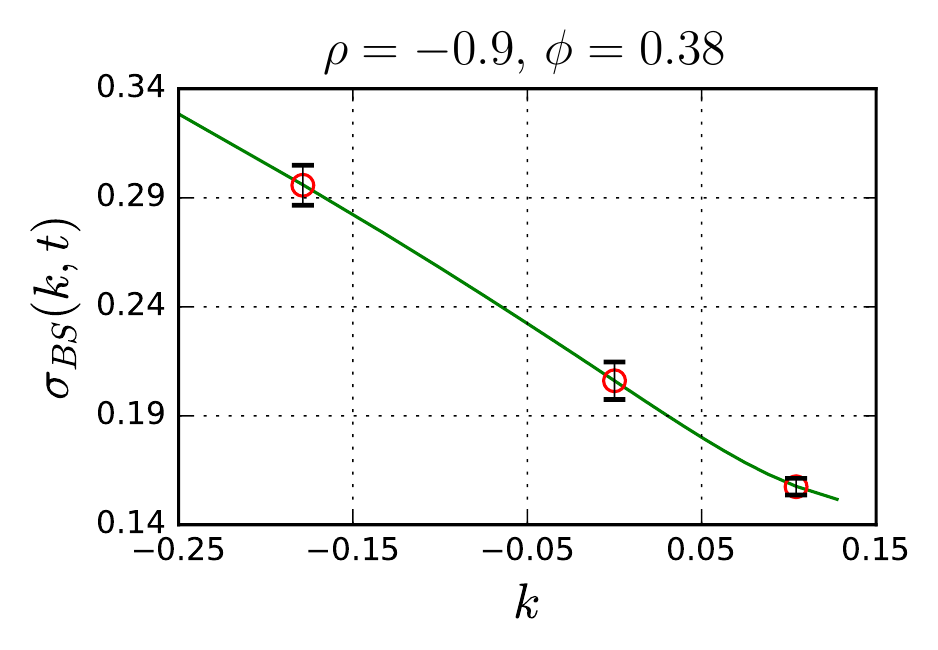}
    \includegraphics[width=0.425\linewidth]{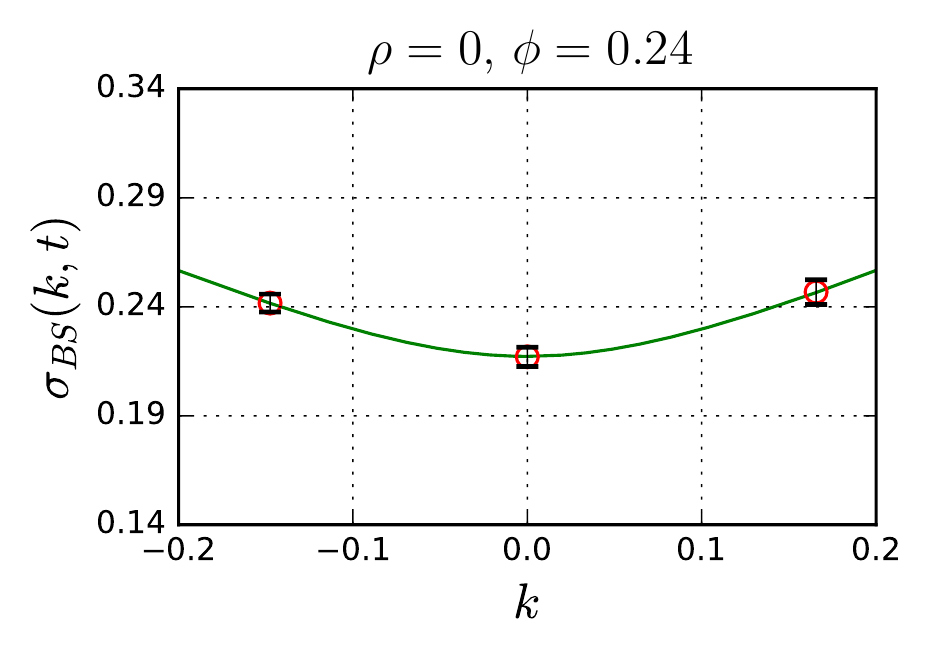}
    \vspace{-5mm}
    \caption{A reproduction of the Base estimator results from Figure \ref{figRMSEsBase}, using instead $n = 8{,}000$ for the case $\rho = -0.9$ and $n = 20{,}500$ for the case $\rho = 0$, in order to match resulting $\phi$ values with those of the Mixed estimator when using $n = 1{,}000$. Ratios of previously observed $\phi^2$ values are used to predict the number of paths required to achieve this.}\label{figRMSEsBase1}
\end{figure}

The relative $\psi^2$ values for the Base and Mixed estimators in Figures \ref{figBaseHist} and \ref{figMixedHist} suggest a 13-fold runtime reduction in the $\rho = -0.9$ regime, and a 34-fold runtime reduction in the $\rho = 0$ regime, in order to match $\phi$ values, thereby a given implied volatility confidence interval. That is, roughly a 20-fold runtime reduction on average. Indeed, in Figure \ref{figRMSEsBase1} we show another set of Base estimator results, which match the Mixed $\phi$ values, requiring $n = 8{,}000$ and $20{,}250$ paths respectively.

Before proceeding, we summarise standard deviations and runtimes for all estimators in Table \ref{tabSummary}, using $n = 1{,}000$. We have no practically meaningful bias to report. To aid a clearer comparison, the Conditional, Controlled and Mixed estimators all utilise antithetic sampling, hence their lower runtimes. The Mixed estimator adopts the variance reducing effects of the Conditional and Controlled estimators in the regimes $\rho = 0$ and $\rho = -1$, respectively. For $-1 < \rho < 0$, the Mixed estimator blends the effects of each, which is already observed in the case of $\rho = -0.9$. Experiment suggests that the Mixed estimator outperforms the Conditional and Controlled estimators best, in a joint sense, around the region $1- \rho^2 = \rho^2$.

\begin{table}[h]
\centering
\begin{tabular}{|l|l|cccc|c|cccc|}
\cline{1-1} \cline{3-6} \cline{8-11}
 &  & \multicolumn{4}{c|}{$\rho = -0.9$} & \multicolumn{1}{l|}{} & \multicolumn{4}{c|}{$\rho = 0$} \\
Estimator &  & 10P & ATM & 10C & $\tau$ & \multicolumn{1}{l|}{} & 10P & ATM & 10C & $\tau$ \\ \cline{1-1} \cline{3-6} \cline{8-11} 
Base &  & 1.28 & 1.24 & 0.52 & 114 &  & 0.94 & 1.03 & 1.25 & 115 \\
Antithetic &  & 1.70 & 1.45 & 0.59 & 49 &  & 0.92 & 0.74 & 1.25 & 49 \\
Conditional &  & 1.19 & 1.02 & 0.34 & 68 &  & 0.26 & 0.15 & 0.28 & 69 \\
Controlled &  & 0.82 & 0.41 & 0.49 & 55 &  & 0.70 & 0.56 & 0.82 & 55 \\
Mixed &  & 0.55 & 0.27 & 0.26 & 71 &  & 0.26 & 0.15 & 0.28 & 70 \\ \cline{1-1} \cline{3-6} \cline{8-11} 
\end{tabular}
\caption{Summary of estimator standard deviations of the samples $\{\hat{\sigma}^{n}_{BS}(k)_i\}_{i=1}^{N}$, using $n = 1{,}000$ paths, with associated average runtimes in milliseconds.}\label{tabSummary}
\end{table}

\subsection{Experiment assessing the accuracy of calibration}

We now briefly demonstrate the impact of these results on an example calibration by simulation of the rBergomi model. We stress that this is only really for illustrative purposes, since knowledge of (untraded) model parameter bounds seems somewhat meaningless without understanding the associated impact on (traded) implied volatility bounds, which we have covered directly. The specification of which rBergomi parameters should be calibrated by simulation is an open question and not a topic we intend to tackle here.

\begin{figure}[!h]
    \centering
    \includegraphics[width=0.425\linewidth]{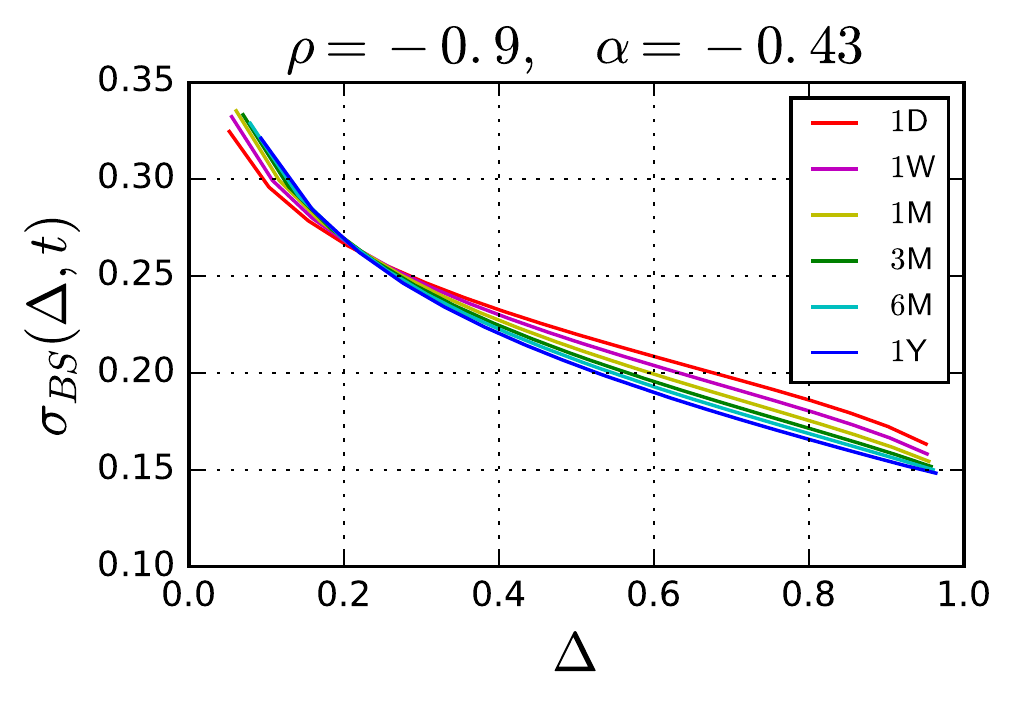}
    \includegraphics[width=0.425\linewidth]{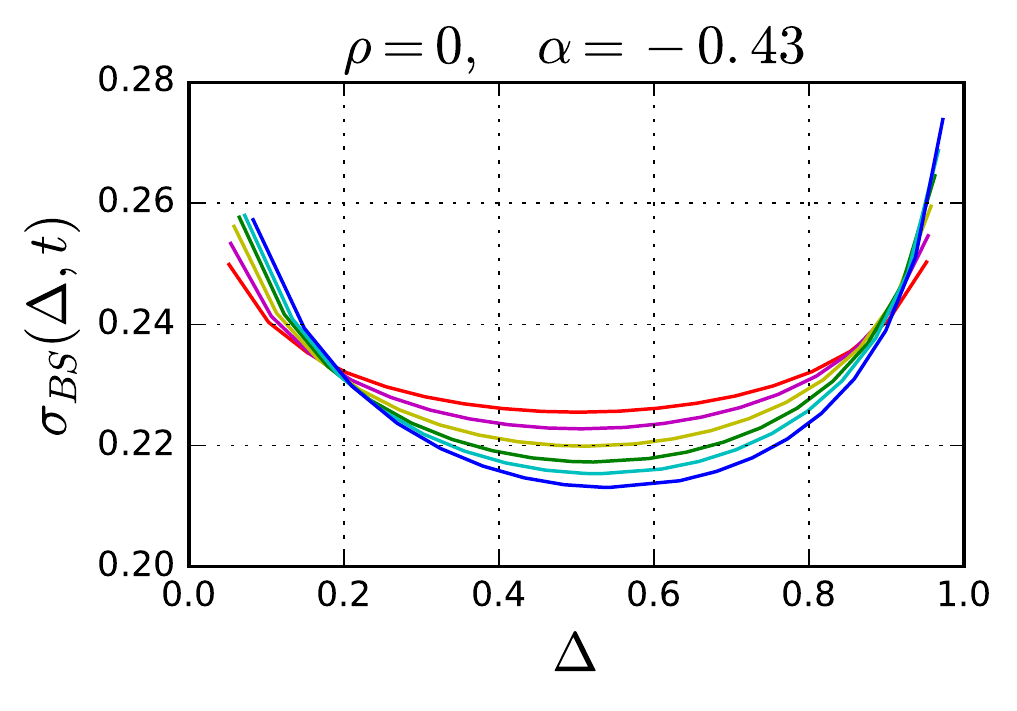}
    \vspace{-5mm}
    \caption{A reproduction of Figure \ref{figSmiles}, in $\Delta$-space, integration of which may lead to $\xi_0(t)$. The near-inhomogeneity across maturities is intriguing. Notice that $\Delta = \mathcal{N}(-d_-)$, referred to as forward delta by \cite{Austing:2014}, is not quite the same as the delta used to define log-strikes, $\mathcal{N}(-d_+)$.}\label{figDelta}
\end{figure}

We assume, to aid this demonstration, that $\alpha$ and $\xi_0(t)$ are fixed by other means at $-0.43$ and $0.235^2$ respectively. This is consistent with the approach adopted by \cite{Jacquier:2017a} for a joint {\small SPX} and {\small VIX} calibration. Therein, $H = \alpha + \frac{1}{2}$ is calibrated pre-simulation to {\small VIX} futures, and $\xi_0(t)$ extracted from an e{\small SSVI} parameterisation \citep{Hendriks:2017} of an observed {\small SPX} implied volatility surface. A more asset class-indifferent approach might be to obtain $\alpha$ from historic time-series using, for example, the methods of \cite{Gatheral:2014} and \cite{Bennedsen:2015}. This is made possible, in theory, since $\alpha$ is preserved in the neat measure change from which the rBergomi model is derived. We suggest a natural approach across asset classes for obtaining $\xi_0(t)$ would be to utilise the elegant integrated variance representation summarised by \cite{Austing:2014},
\begin{equation*}
\int_0^t \xi_0(u) \mathrm{d}u = \mathbb{E}\left[\int_0^t V_u \mathrm{d}u\right] = \int_0^1 \sigma_{BS}(\Delta,t)^2 \mathrm{d}\Delta, \quad \Delta := \mathcal{N}(-d_-),
\end{equation*}
which follows from Fubini's theorem and a change of variables. Clearly this requires an interpolation of observed $\sigma_{BS}(\cdot,t)$ in $\Delta$-space, and some parametric (or piece-wise parametric) assumption for $\xi_0(t)$. We find, however, that even a na\"{i}ve cubic spline across $\sigma_{BS}(\Delta,t)$ and piece-wise constant $\xi_0(t)$ can produce impressive results. In Figure \ref{figDelta}, we reproduce Figure \ref{figSmiles} in $\Delta$-space for the case of $\alpha = -0.43$, given that data sources like Bloomberg do similarly.

\begin{figure}[h]
    \centering
    \begin{tabular}{@{}r@{}r@{}}
    \includegraphics[width=0.425\linewidth]{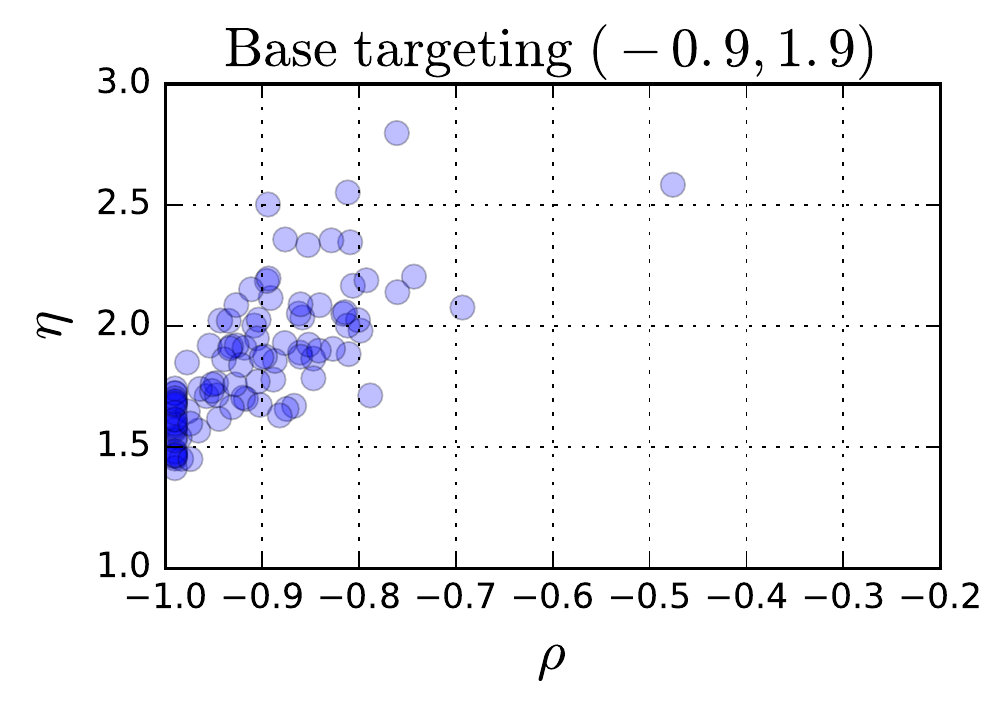} &
    \includegraphics[width=0.425\linewidth]{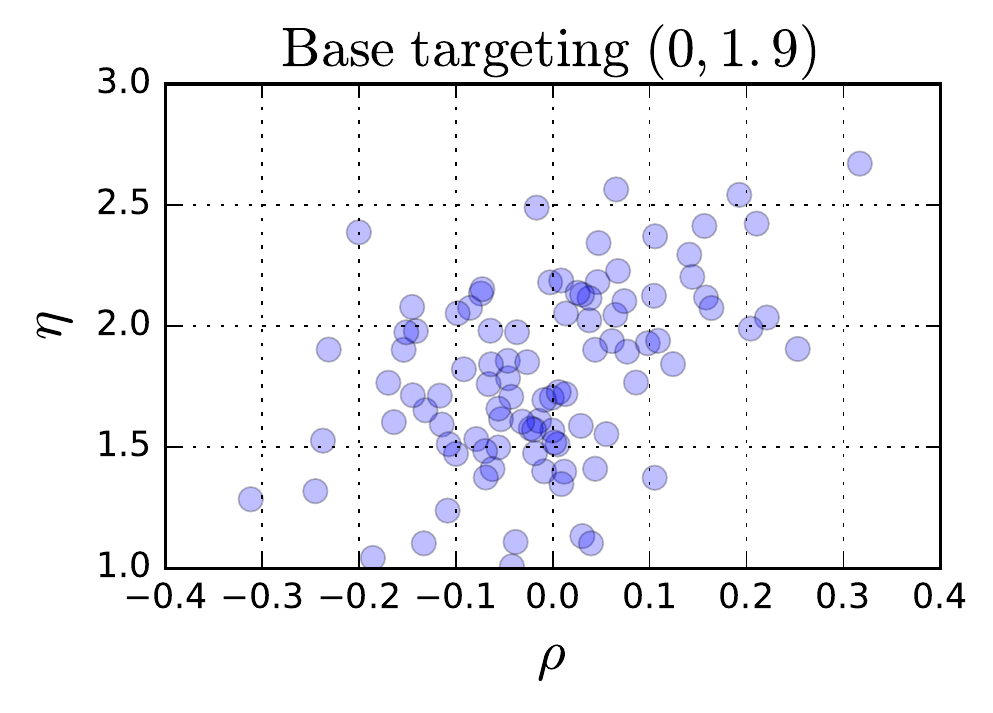} \\
    \includegraphics[width=0.425\linewidth]{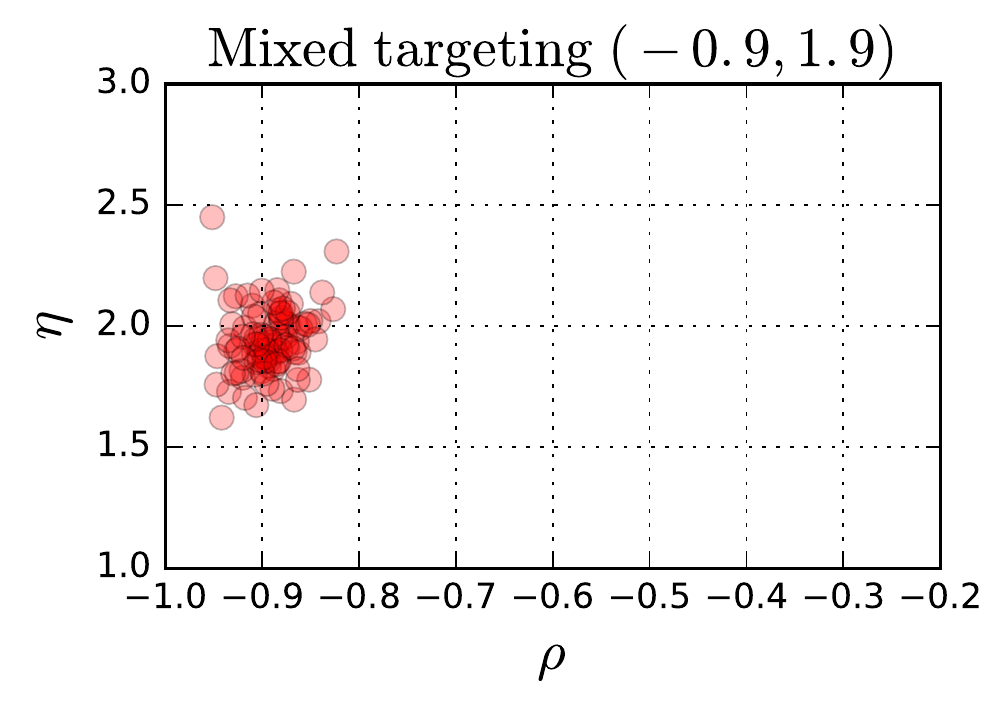} &
    \includegraphics[width=0.425\linewidth]{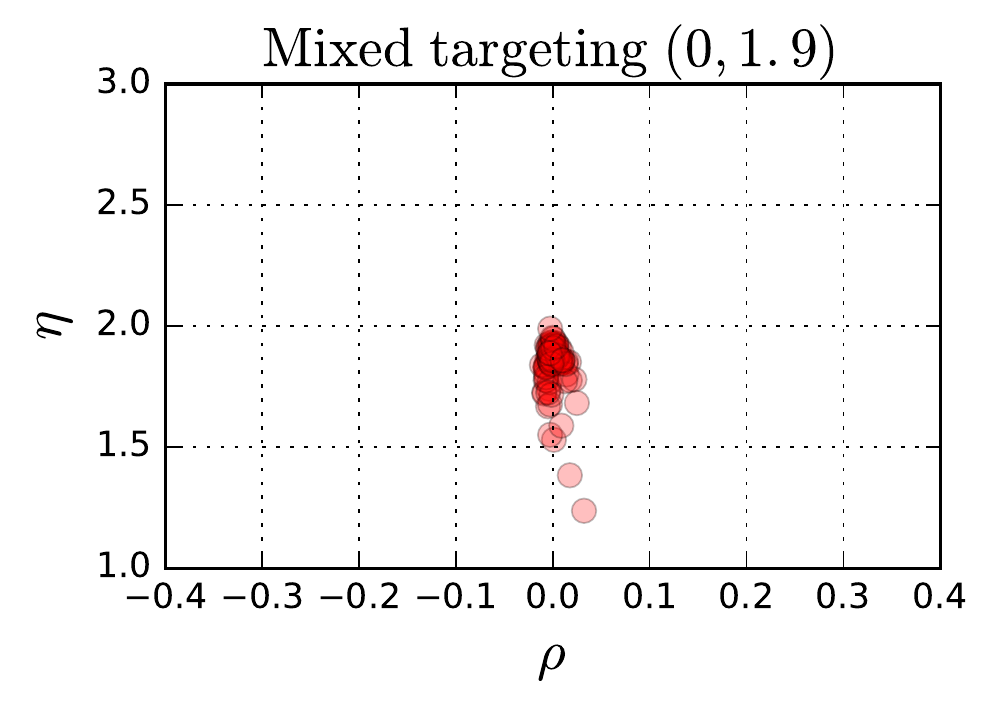}
    \end{tabular}
    \vspace{-5mm}
    \caption{100 calibrations of $\rho$ and $\eta$ using the Base and Mixed estimators, with just $n = 1{,}000$ paths. The minimum of implied volatility absolute {\scriptsize RMSE}s is sought using the {\scriptsize L-BFGS-B} method of \texttt{scipy.optimize.minimize},  with bounds $\rho \in [-0.99,0.99]$, $\eta \in [1.00,3.00]$, allowing this to run for approximately 700 milliseconds. Despite actually making little difference, we initialised the solver for $\rho$ and $\eta$ at the known values in each case, so that the resulting calibrations observed here truly \emph{represent the convergence of $\rho$ and $\eta$ to values away from these known values}---thereby measuring each estimator's \emph{failure} to produce the known distributive properties of the price process, and equivalently the known implied volatilities. The Mixed estimator substantially reduces calibrated $\rho$ and $\eta$ variance, with the Base estimator being somewhat aided in the $\rho = -0.9$ case by the lower bound of $-0.99$.}
    \label{figCalib}
\end{figure}

We proceed to calibrate the rBergomi skew and smile parameters $\rho$ and $\eta$, seeking a minimisation of absolute {\small RMSE}s for the 19 implied volatilites at the {\small 3M} maturity in Figure \ref{figSmiles}. Joint calibrated $\rho$ and $\eta$ distributions are presented in Figure \ref{figCalib}.

\section{Concluding remarks}

We have demonstrated sample paths and the rich implied volatility surfaces generated from the rBergomi model in order to build intuition for its parameters. We have made Python code available on GitHub, from which one is able to replicate these surfaces and generate others. We believe that the potential of rough volatility models is evident and hope that the seeds for practical adoption are now sewn.

Drawing inspiration from \cite{Bergomi:2016}, we have jumped towards the present requirement of rBergomi calibration by simulation, by carefully applying the conditional Monte Carlo method with a control variate and antithetic sampling. Specifically, we have provided a 20-fold runtime reduction on average for achieving a chosen European option implied volatility confidence interval, thus calibration {\small RMSE}.

Although there remain open questions (perhaps most significantly:\ which of the model's parameters, if not all, can be reliably calibrated pre-simulation, and how best?), this is now a thriving area of research in academia, and we are full of resolute optimism. Having practical experience with a variety of stochastic volatility models, we cannot stress enough how central we believe rough processes, like the Volterra process, could be in the future of volatility modelling.

\section*{Acknowledgements}

M.S.P. acknowledges helpful discussions with Chithira Mamallan, who independently obtained results on the effectiveness of antithetic sampling and the Conditional estimator in the context of the rough Bergomi model in her MSci dissertation \citep{Mamallan:2017} at Imperial College London. He also thanks Christian Bayer for discussions on quasi-random numbers.

\end{document}